\documentclass[usenatbib]{mn2e}
\usepackage{graphicx,times}
\usepackage{bm}
\graphicspath{{./fig/}{./png/}}
\newcommand{\EQ}{\begin{equation}}
\newcommand{\EN}{\end{equation}}
\newcommand{\EQA}{\begin{eqnarray}}
\newcommand{\ENA}{\end{eqnarray}}
\newcommand{\eq}[1]{(\ref{#1})}

\newcommand{\Eq}[1]{equation~(\ref{#1})}
\newcommand{\Eqs}[2]{equations~(\ref{#1}) and~(\ref{#2})}

\newcommand{\Sec}[1]{Sect.~\ref{#1}}

\newcommand{\Fig}[1]{Fig.~\ref{#1}}

{}
{}
{}

{}
{}
\newcommand{\meanEMF}{\overline{\mbox{\boldmath ${\cal E}$}}{}}{}
{}
{}
{}
{}
\newcommand{\meanAA}{\overline{\mbox{\boldmath $A$}}{}}{}
\newcommand{\meanBB}{\overline{\mbox{\boldmath $B$}}{}}{}
\newcommand{\meanEE}{\overline{\mbox{\boldmath $E$}}{}}{}
\newcommand{\meanFF}{\overline{\mbox{\boldmath $F$}}{}}{}
\newcommand{\meanFFf}{\overline{\mbox{\boldmath $F$}}_{\rm f}{}}{}
\newcommand{\meanFFm}{\overline{\mbox{\boldmath $F$}}_{\rm m}{}}{}
{}
{}
{}
\newcommand{\meanJJ}{\overline{\mbox{\boldmath $J$}}{}}{}
\newcommand{\meanUU}{\overline{\bm{U}}}
{}
{}
{}
\newcommand{\meanA}{\overline{A}}
\newcommand{\meanB}{\overline{B}}

\newcommand{\meanFm}{\overline{F}_{\rm m}}
\newcommand{\meanFf}{\overline{F}_{\rm f}}
\newcommand{\meanh}{\overline{h}}
\newcommand{\meanhm}{\overline{h}_{\rm m}}
\newcommand{\meanhf}{\overline{h}_{\rm f}}

\newcommand{\meanU}{\overline{U}}

\newcommand{\meanFFF}{\overline{\cal F}}
{}

{}

{}
{}

\newcommand{\alphaK}{\alpha_{\it K}}
\newcommand{\alphaM}{\alpha_{\it M}}
\newcommand{\xxx}{\hat{\mbox{\boldmath $x$}} {}}

\newcommand{\uu}{\mbox{\boldmath $u$} {}}

\newcommand{\bb}{\mbox{\boldmath $b$} {}}
\newcommand{\BB}{\mbox{\boldmath $B$} {}}

\newcommand{\jj}{\mbox{\boldmath $j$} {}}
\newcommand{\JJ}{\mbox{\boldmath $J$} {}}

\newcommand{\nab}{\mbox{\boldmath $\nabla$} {}}

\newcommand{\oo}{\mbox{\boldmath $\omega$} {}}

\newcommand{\dd}{{\rm d} {}}
\newcommand{\const}{{\rm const}  {}}

\def\Rm{R_{\rm m}}

\def\kf{k_{\rm f}}

\def\Hm{H_{\rm m}}

\def\etat{\eta_{\rm t}}

\def\Beq{B_{\rm eq}}

\def\half{{\textstyle{1\over2}}}

\def\onethird{{\textstyle{1\over3}}}

\newcommand{\yjgr}[3]{ #1, {J.\ Geophys.\ Res.,} {#2}, #3}

\newcommand{\yapj}[3]{ #1, {ApJ,} {#2}, #3}

\newcommand{\yapjl}[3]{ #1, {ApJ,} {#2}, #3}

\newcommand{\yan}[3]{ #1, {Astron.\ Nachr.,} {#2}, #3}

\newcommand{\yana}[3]{ #1, {A\&A,} {#2}, #3}

\newcommand{\yjfm}[3]{ #1, {J.\ Fluid Mech.,} {#2}, #3}
\newcommand{\ypepi}[3]{ #1, {Phys.\ Earth Planet.\ Int.,} {#2}, #3}

\newcommand{\ypp}[3]{ #1, {Phys.\ Plasmas,} {#2}, #3}

\newcommand{\yprl}[3]{ #1, {Phys.\ Rev.\ Lett.,} {#2}, #3}

\newcommand{\ymn}[3]{ #1, {MNRAS,} {#2}, #3}

\newcommand{\ypre}[3]{ #1, {Phys.\ Rev.\ E,} {#2}, #3}

\newcommand{\yjour}[4]{ #1, {#2}, {#3}, #4}

\newcommand{\ybook}[3]{ #1, {#2} (#3)}
\newcommand{\yproc}[5]{ #1, in {#3}, ed.\ #4 (#5), #2}

\newcommand{\papj}[1]{ #1, {ApJ}, to be published}

\newcommand{\sprl}[1]{ #1, {PRL}, submitted}

\title[Magnetic helicity losses from a mean-field dynamo]
{Small-scale magnetic helicity losses from a mean-field dynamo}
\author[
]{
Axel Brandenburg$^1$, Simon Candelaresi$^1$ and Piyali Chatterjee$^2$\\
$^1$NORDITA, AlbaNova University Center, Roslagstullsbacken 23,
SE-10691 Stockholm, Sweden\\
$^2$Department of Astronomy and Astrophysics,
Tata Institute of Fundamental Research, Colaba, Mumbai 400005, India}

\date{\today,~ $ $Revision: 1.36 $ $}
\begin{document}

\maketitle

\begin{abstract}
Using mean-field models with a dynamical quenching formalism we show that
in finite domains magnetic helicity fluxes associated with small-scale
magnetic fields are able to alleviate catastrophic quenching.
We consider fluxes that result either from advection by a mean flow,
the turbulent mixing down the gradient of mean small-scale magnetic
helicity concentration, or the explicit removal which may be associated
with the effects of coronal mass ejections in the Sun.
In the absence of shear, all the small-scale magnetic helicity fluxes are
found to be equally strong both for large-scale and small-scale fields.
In the presence of shear there is also an additional magnetic
helicity flux associated with the mean field, but this flux does not
alleviate catastrophic quenching.
Outside the dynamo-active region there are neither sources nor sinks of
magnetic helicity, so in a steady state this flux must be constant.
It is shown that unphysical behavior emerges if the small-scale magnetic
helicity flux is forced to vanish within the computational domain.
\end{abstract}
\label{firstpage}
\begin{keywords}
magnetic fields --- MHD --- hydrodynamics -- turbulence
\end{keywords}

\section{Introduction}

Both mean-field theories as well as direct simulations
of the generation of large-scale magnetic fields
in astrophysical bodies such as the Sun or the Galaxy invoke
the effects of twist.
Twist is typically the result of the Coriolis force acting on
ascending or descending magnetic field structures in a stratified medium.
The net effect of this systematic twisting motion on the magnetic field
is called the $\alpha$ effect.
In text books the $\alpha$ effect is normally introduced as a result
of helical turbulence \citep{Mof78,Par79,KR80}, but it could also arise
from magnetic buoyancy instabilities \citep{Sch87,BS98}.
The latter may also be at the heart of what is known as the
Babcock--Leighton mechanism that describes the net effect of the tilt
of decaying active regions.
Mathematically, this mechanism
can also be described by an $\alpha$ effect \citep{Stix74}.
Regardless of all these details, any of these processes face a serious
challenge connected with the conservation of magnetic helicity
\citep{PFL76,KR82,KRR95}.
The seriousness of this is not generally appreciated, even though the
conservation of magnetic helicity has long been associated with what is called
catastrophic $\alpha$ quenching \citep{GD94,GD95,GD96}.
Catastrophic $\alpha$ quenching refers to the fact that the $\alpha$ effect
in helical turbulence in a periodic box decreases with increasing magnetic
Reynolds number for equipartition strength magnetic fields \citep{VC92,CH96}.
This would be `catastrophic' because the magnetic Reynolds number is large
($10^9$ in the Sun and $10^{15}$ in the Galaxy).

A promising theory for modeling catastrophic $\alpha$ quenching in a
mean-field simulation is the dynamical quenching formula,
i.e.\ an evolution equation for the $\alpha$ effect that follows from
magnetic helicity conservation \citep{KR82}.
Later, \cite{FB02} showed for the first time that this formalism is also able
to describe the slow saturation of a helical dynamo in a triply-periodic
domain \citep{B01}.
As this dynamo runs into saturation, a large-scale magnetic field builds
up, but this field possesses magnetic helicity.
Indeed, the eigenfunction of a homogeneous $\alpha^2$ dynamo has
magnetic and current helicities proportional to $\alpha$.
However, this concerns only the mean field, and since the helicity of
the total field is conserved, the small-scale or fluctuating field must
have magnetic helicity of the opposite sign \citep{See96}.
This leads to a reduction of the $\alpha$ effect \citep{PFL76}.

The dynamical quenching formalism is now frequently used to model
the nonlinear behavior of mean-field dynamos with and without shear
\citep{BB02}, open or closed boundaries \citep{BS05}, and sometimes
even without $\alpha$ effect \citep{YBR03,BS05}.
However, it became soon clear that the catastrophic quenching of the
$\alpha$ effect can only be alleviated in the presence of magnetic
helicity fluxes out of the domain \citep{BF00a,BF00b,KMRS00,KMRS02}.
There are various contributions to the magnetic helicity flux
\citep{RK00,VC01,SB04,SB06}, but one of the most obvious ones
is that associated with advection.
\cite{SSSB06} have implemented this effect in a mean-field model
with dynamical quenching in order to model the effects of a
wind on the evolution of the galactic magnetic field.
One goal of the present paper is to study this effect in more detail.
In particular, it is important to clarify the consequences of boundary
conditions on the local dynamics away from the boundaries.
Indeed, is it really true that a helicity flux has to be
maintained all the way to the boundaries, or can the helicity
flux be confined to part of the domain to alleviate catastrophic
$\alpha$ quenching at least locally?
What happens if this is not the case?

The notion of alleviating catastrophic $\alpha$ quenching only locally
is sometimes invoked in models of the solar dynamo that rely on the
production of strong magnetic fields at the bottom of the convection zone.
By placing the $\alpha$ effect only near the surface, as is done
in the interface dynamo of \cite{Par93} or dynamos that are controlled
by meridional circulation \citep{CSD95}, one may evade catastrophic
quenching more easily.
On the other hand, as shown by \cite{YBR03}, the effects of magnetic
helicity conservation can even play a role even if there is originally
no $\alpha$ effect.
It is therefore important to understand in more detail the physics
of dynamical $\alpha$ quenching and its dependence on magnetic helicity
fluxes.

Our starting point in this paper is the model of \cite{SSSB06}, where
magnetic helicity fluxes were driven by the advection from a wind.
This allows us to study the effects of varying strength of this flux
in different parts of the domain.
For simplicity, and in order to isolate the main effects,
we ignore shear in most parts of this paper.
In view of later applications to the Sun and the Galaxy this is clearly
artificial, but it helps significantly in the interpretation of the results.
In particular, in the absence of shear, it is possible to have steady
solutions, or at least solutions whose magnetic energy density is
constant in time.
This simplifies the interpretation of the results.

\section{The model}

\subsection{Evolution equation of the mean-field}

In this paper we consider a simple mean-field dynamo in a local
one-dimensional domain.
Such a model could be applicable to one hemisphere of a rotating
disc or to the region close to the equator of outer stellar
convection zones.
Denoting the mean magnetic field by $\meanBB=\meanBB(z,t)$,
the coordinate $z$ would correspond either to the height above the
midplane in the case of the disc, or to the latitudinal distance from
the equator in the case of a spherical shell.
The $x$ and $y$ components would correspond to poloidal and toroidal
fields, although in the absence of shear the two are interchangeable
and cannot be distinguished.
Using $\nab\cdot\meanBB=\partial\meanB_z/\partial z=0$, we have
$\meanB_z=\const=0$, i.e.\ no $\meanB_z$ field is imposed.
Such a mean field could be obtained by averaging the actual magnetic
field over the $x$ and $y$ directions on a Cartesian domain.

The evolution of $\meanBB$ is governed by the Faraday equation
\EQ
{\partial\meanBB\over\partial t}=-\nab\times\meanEE,
\label{dBbar}
\EN
where
$\meanEE=-(\meanUU_{\rm S}+\meanUU)\times\meanBB-\meanEMF+\eta\mu_0\meanJJ$
is the mean electric field, $\meanUU$ is the mean flow in the $z$ direction,
$\meanUU_{\rm S}=(0,Sx,0)$ is a linear shear flow,
$\meanEMF$ is the mean electromotive force,
$\meanJJ=\nab\times\meanBB/\mu_0$ is the mean current density,
and $\mu_0$ is the vacuum permeability.
In one case we adopt a shear parameter $S$ that is different from zero.
Since the shear is linear, we can write
$\meanUU_{\rm S}\times\meanBB$ as $-S\meanA_y\xxx$ plus a gradient
term that can be removed by a gauge transformation.
Thus, we have
\EQ
-\meanEE=\nab(Sx\meanA_y)-S\meanA_y\xxx
+\meanUU\times\meanBB+\meanEMF-\eta\mu_0\meanJJ,
\label{meanEE2}
\EN
where $\meanUU$ is now the flow associated with the wind only
and does not include the shear flow.
Next, we express $\meanBB=\nab\times\meanAA$ in terms of the magnetic vector
potential $\meanAA$, and solve \Eq{dBbar} in its uncurled form,
$\partial\meanAA/\partial t=-\meanEE-\nab\overline\phi$,
where $\overline\phi$ is the mean electrostatic potential.
We perform a gauge transformation, $\meanAA\to\meanAA+\nab\Lambda$,
with the choice $\Lambda=\int(\overline\phi-Sx\meanA_y)\,\dd t$, which
removes the gradient term to yield
\EQ
{\partial\meanAA\over\partial t}=-\meanEE,
\label{dAbar}
\EN
which is then the final form of our equation for $\meanAA$.
This form of the equation together with boundary conditions for $\meanAA$
characterize the gauge used to calculate magnetic helicity densities
and magnetic helicity fluxes for the mean field.

We solve \Eq{dAbar} in the domain $0<z<L$ and assume either a
vacuum or a perfect conductor boundary condition on $z=L$.
This means that on $z=L$ the mean magnetic field either vanishes,
i.e.\ $\meanB_x=\meanB_y=0$, or that its $z$ derivative
vanishes, i.e.\ $\meanB_{x,z}=\meanB_{y,z}=0$, where a comma
denotes partial differentiation.
In terms of $\meanAA$ this means that on $z=L$ we have either
\EQ
\meanA_{x,z}=\meanA_{y,z}=0\quad\mbox{(vacuum condition)},
\EN
or
\EQ
\meanA_{x}=\meanA_{y}=0\quad\mbox{(perfect conductor condition)}.
\label{percond}
\EN
It is well known that the solutions can be in one of two pure parity
states that are either symmetric (S) or antisymmetric (A) about the midplane
\citep{KR80}, so we have either $\meanB_{x,z}=\meanB_{y,z}=0$ or
$\meanB_x=\meanB_y=0$ on $z=0$.
In terms of $\meanAA$ this means either
\EQ
\meanA_x=\meanA_y=0\quad\mbox{on $z=0$}\quad\mbox{(S solution)}
\label{Ssolution}
\EN
or
\EQ
\meanA_{x,z}=\meanA_{y,z}=0\quad\mbox{on $z=0$}\quad\mbox{(A solution)}.
\EN
We note that the particular boundary conditions \eq{percond} and
\eq{Ssolution} fix the value of $\meanAA$ on $z=L$ or $z=0$, respectively.
In all other combinations the value of $\meanAA$ is not fixed and the
magnetic helicity could exhibit an unphysical drift \citep{BDS02}.
However, in the present paper we study magnetic helicity density and
its flux only in situations where either \eq{percond} or \eq{Ssolution}
are used.

We recall that, even though there is no $\Omega$ effect, i.e.\ no mean
flow in the $y$ direction, we shall allow for a flow $\meanUU$ in the
$z$ direction.
In a disc this would correspond to a vertical wind, while in a star this
might locally be associated with meridional circulation.

\subsection{Magnetic helicity conservation}

In this paper we will study the evolution of magnetic helicity of
mean and fluctuating fields.
In our gauge, the evolution of the magnetic helicity density of the
mean field, $\Hm=\meanAA\cdot\meanBB$, is given by
\EQ
{\partial\meanhm\over\partial t}=
2\meanEMF\cdot\meanBB-2\eta\mu_0\meanJJ\cdot\meanBB
-\nab\cdot\meanFFm,
\label{dhm}
\EN
where $\meanFFm=\meanEE\times\meanAA$ is the flux of magnetic helicity
of the mean magnetic field.
Under the assumption of scale separation, \cite{SB06} have defined a
magnetic helicity density of the small-scale field in terms of its
mutual linkages.
They derived an evolution equation for the magnetic helicity density of
the small-scale field,
\EQ
{\partial\meanhf\over\partial t}=
-2\meanEMF\cdot\meanBB-2\eta\mu_0\overline{\jj\cdot\bb}
-\nab\cdot\meanFFf,
\label{dhf}
\EN
which is similar to \Eq{dhm}, except that the $\meanEMF\cdot\meanBB$
appears with the opposite sign.
This implies that turbulent amplification and diffusion of mean magnetic
field (characterized by the $\meanEMF$ term)
cannot change the total magnetic helicity density, $\meanh=\meanhm+\meanhf$,
which therefore obeys the equation
\EQ
{\partial\meanh\over\partial t}=
-2\eta\mu_0\overline{\JJ\cdot\BB}
-\nab\cdot\meanFF,
\label{dh}
\EN
where $\meanFF=\meanFFm+\meanFFf$ is the total magnetic helicity flux,
and $\overline{\JJ\cdot\BB}=\meanJJ\cdot\meanBB+\overline{\jj\cdot\bb}$
is the total current helicity density.

\subsection{Dynamical quenching formalism}

In order to satisfy the evolution equation for the total magnetic helicity
density \eq{dh}, we have to solve \Eq{dhf} along with \Eq{dAbar}, which implies
that \Eqs{dhm}{dh} are automatically obeyed.
We make the assumption that the turbulence is at small scales nearly
isotropic.
This means that $\mu_0\overline{\jj\cdot\bb}\approx\kf^2\meanhf$.
The $\overline{\jj\cdot\bb}$ term also modifies the mean electromotive
force by producing an $\alpha$ effect \citep{PFL76}.
This is sometimes referred to as the magnetic $\alpha$ effect,
\EQ
\alphaM=\onethird\tau\overline{\jj\cdot\bb}/\overline\rho,
\EN
where $\tau$ is the correlation time of the turbulence.
In the following we ignore compressibility effects and assume that
the mean density $\overline\rho$ is constant\footnote{Note that a
constant mean density implies that there must exist a small-scale
mass flux compensating the losses associated with the mass flux
$\overline\rho\meanUU$.}
Next, we assume that the turbulence is helical, so there is also a
kinetic $\alpha$ effect proportional to the kinetic helicity,
\EQ
\alphaK=-\onethird\tau\overline{\oo\cdot\uu},
\EN
where $\oo=\nab\times\uu$ is the vorticity.
The total $\alpha$ effect is then
\EQ
\alpha=\alphaK+\alphaM,
\label{alptot}
\EN
and the resulting mean electromotive force is
\EQ
\meanEMF=\alpha\meanBB-\etat\mu_0\meanJJ,
\EN
where
\EQ
\etat=\onethird\tau\overline{\uu^2}
\EN
is the turbulent magnetic diffusivity.
In the following we consider $\etat$ and $\eta$ as given and
define their ratio as the magnetic Reynolds number, $\Rm=\etat/\eta$.
We shall express the strength of the magnetic field in terms
of the equipartition value,
\EQ
\Beq=(\mu_0\overline\rho\overline{\uu^2})^{1/2},
\EN
which allows us to determine $\tau$ in the mean-field model via
$\onethird\tau=\mu_0\overline\rho\etat/\Beq^2$.
We characterize the value of $\eta$ in terms of the magnetic Reynolds
number, $\Rm=\etat/\eta$.
With these preparations we can write the dynamical
quenching formula as
\EQ
{\partial\alphaM\over\partial t}=
-2\etat\kf^2\left({\meanEMF\cdot\meanBB\over\Beq^2}+{\alphaM\over\Rm}\right)
-{\partial\over\partial z}\meanFFF_\alpha,
\label{dalpMdt}
\EN
where $\meanFFF_\alpha$ is related to the mean magnetic
helicity flux of the fluctuating field via
\EQ
\meanFFF_\alpha={\mu_0\overline\rho\etat \kf^2 \over\Beq^2}\,\meanFFf.
\label{Falp}
\EN
In order to compute mean-field models we have to solve \Eq{dAbar} together
with \Eq{dalpMdt} using a closed expression for the flux $\meanFFF_\alpha$.
In this paper we focus on the advective flux proportional to $\alphaM\meanUU$,
but in some cases we consider instead the effects of a turbulent
magnetic helicity flux that we model by a Fickian diffusion term
proportional to $-\kappa_\alpha\nab\alphaM$,
where $\kappa_\alpha$ is a diffusion term that is either zero or otherwise
a small fraction of $\etat$.

In addition, we consider cases where we model magnetic helicity fluxes
by an explicit removal of $\meanhf$ from the domain in regular time
intervals $\Delta t$.
Such an explicit removal of magnetic helicity associated with the
fluctuating field may model the effects of coronal mass ejections,
although one would expect that in reality such an approach also
implies some loss of magnetic helicity associated with the large-scale field.
The approach of removing the fluctuating magnetic field was
employed by \cite{BDS02} in connection with three-dimensional
turbulence simulations to demonstrate that it is, at least in principle,
possible to alleviate catastrophic quenching by an artificial filtering
out of small-scale turbulent magnetic fields.
In the present paper we model the occasional removal of $\meanhf$
by resetting its values
\EQ
\meanhf\to\meanhf-\Delta\meanhf\quad\mbox{in regular intervals $\Delta t$},
\label{removal}
\EN
where $\Delta\meanhf=\epsilon\meanhf$ is chosen to be a certain fraction
$\epsilon$ of the current value of $\meanhf$.
In our one-dimensional model the corresponding expression for the flux
$\Delta\meanFf$ can be obtained by integration, i.e.\
\EQ
\Delta\meanFf(z,t)=\int_0^z\Delta\meanhf(z',t)\,\dd z'.
\EN
Since magnetic helicity densities and their fluxes are proportional to
each other, we have simply
\EQ
\meanFFF_\alpha=\alphaM\meanU-\kappa_\alpha{\partial\alphaM\over\partial z}
+\Delta\meanFFF_\alpha,
\label{FluxFormula}
\EN
where
$\Delta\meanFFF_\alpha=(\mu_0\overline\rho\etat \kf^2/\Beq^2)\Delta\meanFf$
is defined analogously to \Eq{Falp}.

We note that the $\alpha$ effect will produce magnetic fields that
have magnetic helicity with the same sign as that of $\alpha$,
and the rate of magnetic helicity production is proportional to
$\alpha\meanBB^2$.
In the northern hemisphere we have $\alpha>0$, so the mean
field should have positive magnetic helicity.
We recall that shear does not contribute to magnetic helicity production,
because the negative electric field associated with the shear flow,
$\meanUU_{\rm S}\times\meanBB$, gives no contribution to magnetic
helicity production, which is proportional to $\meanEE\cdot\meanBB$,
but it can still give a contribution to the flux of magnetic helicity.
This is also evident if we write shear using the $-S\meanA_y\xxx$ term
in \Eq{meanEE2}: after multiplying with $\meanBB$ and using
$\meanB_x=\partial\meanA_y/\partial z$, we find that this term can be
integrated to give just an additional flux term, $\half S\meanA_y^2$.
However, this contribution belongs clearly to the magnetic helicity flux
associated with the large-scale field and is therefore unable to alleviate
catastrophic quenching.

\subsection{Model profiles and boundary conditions}

\begin{figure}\begin{center}
\includegraphics[width=\columnwidth]{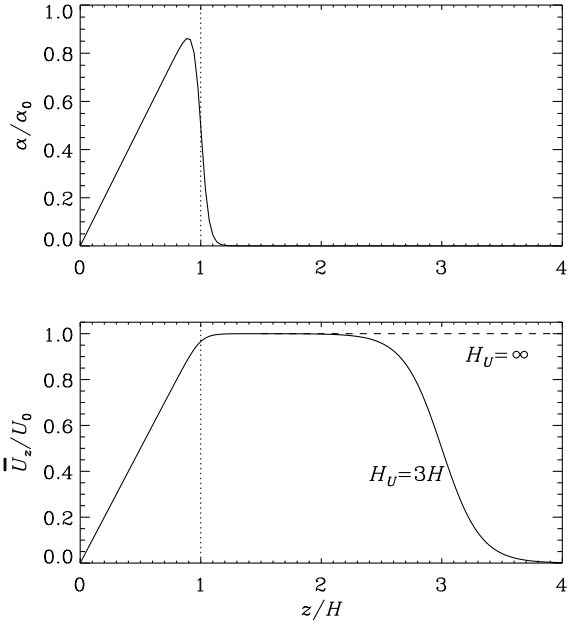}
\end{center}\caption[]{
Profiles of $\alpha$ and $\meanU$
for $w_\alpha k_1=0.2$ and $w_Uk_1=1$.
}\label{pprof}\end{figure}

We consider a model similar to that of \cite{SSSB06} who adopted
linear profiles for $\alphaK$ and $\meanUU$ of the form
$\alphaK=\alpha_0z/H$ and $\meanU_z=U_0z/H$, where the height $H$
was chosen to be equal to the domain size, $H=L$.
However, in order to separate boundary effects from effects of the dynamo
we also consider the case where we
extend the domain in the $z$ direction and choose $L=4H$ and let $\alphaK$ go
smoothly to zero at $z=H$ and $\meanU_z$ either goes to a constant for
$z>H$ or it also goes smoothly to zero.
Thus, we choose
\EQ
\alpha=\alpha_0{z\over H}\Theta(z;H,w_\alpha),
\EN
where we have defined the profile function
\EQ
\Theta(z;H,w)=\half\left(1-\tanh{z-H\over w}\right),
\EN
which is unity for $z\ll H$ and zero otherwise,
and $w$ quantifies the width of this transition.
For the wind we choose the function
\EQ
\meanU_z=U_0{z\over H}\left[1+(z/H)^n\right]^{-1/n}
\Theta(z;H_U,w_U),
\EN
with $n=20$.
Both profiles are shown in \Fig{pprof}.
The strictly linear profiles of \cite{SSSB06} can be recovered by taking $L=H$,
$w_\alpha\to0$, and $n\to\infty$.

As length unit we take $k_1=\pi/2H$, and as time unit
we take $(\etat k_1^2)^{-1}$.
This deviates from \cite{SSSB06}, who used $\pi/H$ as their
basic wavenumber.
Our motivation for this change is that now the
turbulent decay rate is equal to $\etat k_1^2$,
without an extra 1/4 factor.
We adopt nondimensional measures
for $\alpha_0$, $U_0$, and $S$, by defining
\EQ
C_\alpha={\alpha_0\over\etat k_1},\quad
C_U={U_0\over\etat k_1},\quad\mbox{and}\quad
C_S={S\over\etat k_1^2}.
\EN
To match the parameters of \cite{SSSB06}, we note that
$C_U=0.6$ corresponds to their value of 0.3,
and the value $\kf/k_1=10$ corresponds to their value of 5.

We obtain solutions numerically using two different codes.
One code uses an explicit third-order Runge-Kutta time stepping
scheme and the other one a semi-implicit scheme.
Both schemes employ a second order finite differences.
We begin by reporting results for the original profile of
\cite{SSSB06} with $L=H$.

\section{Results}

\subsection{Kinematic behavior of the solutions}
\label{Kinematic}

\begin{figure}\begin{center}
\includegraphics[width=\columnwidth]{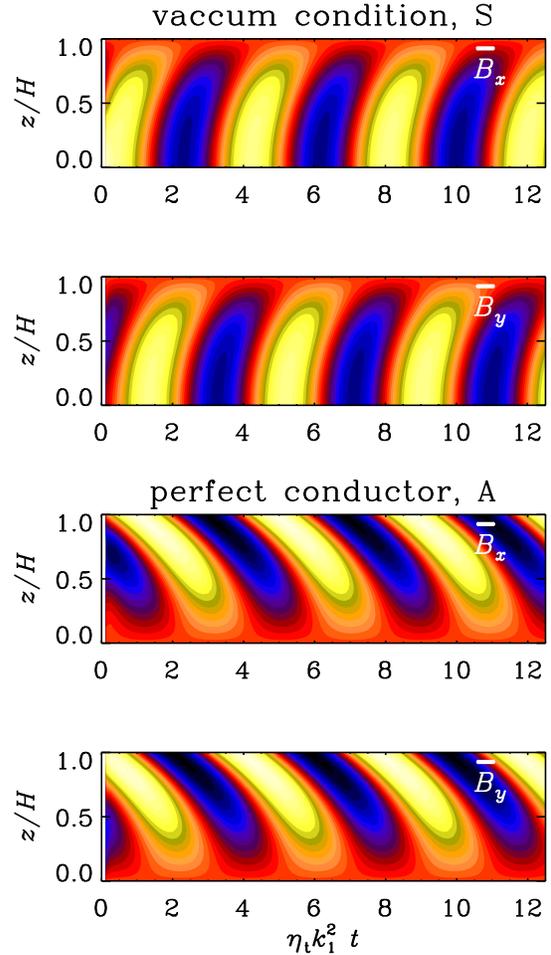}
\end{center}\caption[]{
Space-time diagrams for $\meanB_x$ and $\meanB_y$ for
the marginal values of $C_\alpha$ for $L=H$ with $C_U=0$
and either the symmetric solution (S) with a vacuum boundary condition on $z=H$
or the antisymmetric solution (A) with the perfect conductor boundary condition.
In both cases the critical value $C_\alpha=5.13$ is applied.
Light (yellow) shades indicate positive values and dark (blue) shades
indicate negative values.
}\label{pbutter_kin}\end{figure}

When the magnetic field is weak, the backreaction via the Lorentz force
and hence the $\alphaM$ term are negligible.
The value of $\Rm$ does then not enter into the theory.
The effects of magnetic helicity fluxes are therefore not important,
so we begin by neglecting the wind or other transporters of magnetic helicity.
For the linear $\alpha$ profile we find that the critical value of
$C_\alpha$ for dynamo action to occur is about 5.13.
These solutions are oscillatory with a dimensionless frequency
$\tilde{\omega}\equiv\omega/\etat k_1^2=1.64$.
The oscillations are associated with a migration in the positive $z$
direction.
This is shown in \Fig{pbutter_kin} where we compare with the case of
a perfectly conducting boundary condition at $z=H$ for which we find
$C_\alpha^{\rm crit}=7.12$ and $\tilde{\omega}=2.28$.

The fact that there are oscillatory solutions to the $\alpha^2$ dynamo is
perhaps somewhat unusual, but it is here related to the fact that $\alpha$
changes sign about the equator.
Similar behavior has been seen in some other $\alpha^2$ dynamos where
$\alpha$ changes sign with depth \citep{SG03,REO03,RH04,GZR05} and in
simulations of helically forced turbulence with a change of sign
about the equator \citep{Dhruba}.
In the latter case, however, the outer boundaries were perfectly conducting.
In our mean-field model such a case is also oscillatory, as will be
discussed below.

Note that we have made here the assumption that the solutions are
symmetric about the midplane, i.e.\ $\meanB_i(z,t)=\meanB_i(-z,t)$
for $i=x$ or $y$.
For the application to real systems such a symmetry condition
can only be justified if the symmetric solution is more easily
excited than the antisymmetric one for which $\meanB_i(z,t)=-\meanB_i(-z,t)$
for $i=x$ or $y$.
This is indeed the case when we adopt the vacuum condition at
$z=H$, because the antisymmetric solution has $C_\alpha^{\rm crit}=7.14$
in that case.
However, this is not the case for the perfect conductor
boundary condition for which the antisymmetric solution has
$C_\alpha^{\rm crit}=5.12$.
We remark that there is a striking correspondence in the critical $C_\alpha$
values between the antisymmetric solution with perfect conductor boundary
condition and the symmetric solution with vacuum condition on the one hand,
and the symmetric solution with perfect conductor condition and the
antisymmetric solution with vacuum condition on the other hand.

In the following we consider both symmetric solutions using the
vacuum boundary conditions, as well as antisymmetric ones using the
perfect conductor boundary condition, which corresponds in each case
to the most easily excited mode.
In the cases where we use a vacuum condition we shall sometimes also
apply a wind.
This makes the dynamo somewhat harder to excite and
raises $C_\alpha^{\rm crit}$ from 5.12 to 5.60 for $C_U=0.6$,
but the associated magnetic helicity flux alleviates catastrophic
quenching in the nonlinear case.
Alternatively, we consider an explicit removal of magnetic helicity
to alleviate catastrophic quenching.
In cases with perfect conductor boundary conditions the most easily
excited mode is antisymmetric about the equator, which corresponds
to a boundary condition that permits a magnetic helicity flux through
the equator.
This would not be the case for the symmetric solutions.

\subsection{Saturation behavior for different values of $\Rm$}

We now consider the saturated state for a value of $C_\alpha$
that is supercritical for dynamo action.
In the following we choose the value $C_\alpha=8$.
The saturation behavior is governed by \Eq{dalpMdt}.
Throughout this paper we assume $\kf/k_1=10$ for the scale separation ratio.
This corresponds to the value 5 in \cite{SSSB06}, where $k_1$ was
defined differently.
The dynamo saturates by building up negative $\alphaM$ when $\alphaK$
is positive.
This diminishes the total $\alpha$ in \Eq{alptot} and saturates the dynamo.
The strength of this quenching can be alleviated by magnetic helicity
fluxes that lower the negative value of $\alphaM$.

\begin{figure}\begin{center}
\includegraphics[width=\columnwidth]{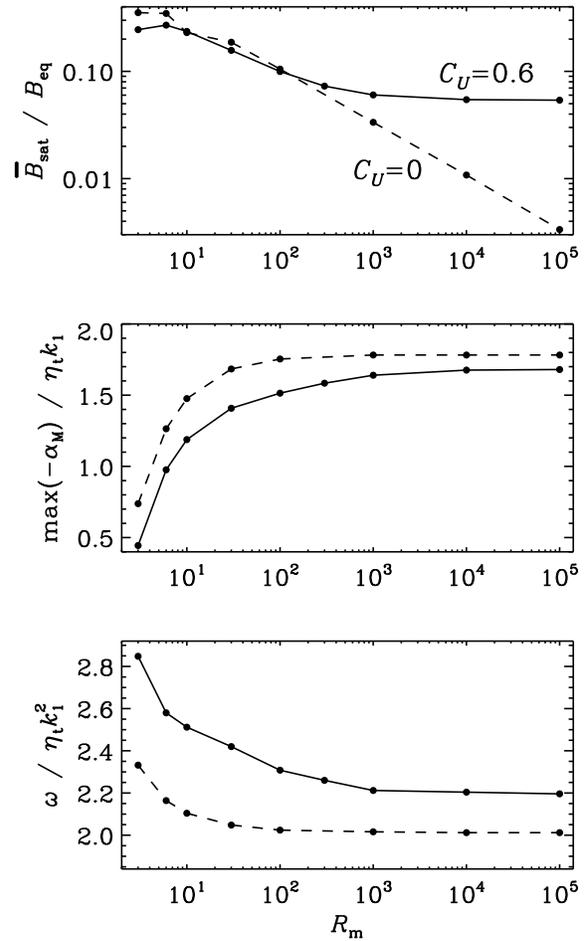}
\end{center}\caption[]{
Scaling of the extremal value of $\alphaM$,
the saturation field strength $\meanB_{\rm sat}$,
and the cycle frequency $\omega$ with $\Rm$
and either $C_U=0.6$ (solid lines) or $C_U=0$ (dashed lines).
}\label{pscaling}\end{figure}

We plot in \Fig{pscaling} the dependence of the saturation field strength
$\meanB_{\rm sat}$, defined here as the maximum of $|\meanBB(z)|$ at the
time of saturation.
To monitor the degree of quenching we also plot in \Fig{pscaling} the
$\Rm$ dependence of the maximum of the negative value of $\alphaM$ at
the time when the dynamo has saturated and reached a steady state.
The maximum value of $-\alphaM$ is lowered by about 5\% from 1.8 to 1.7
in units of $\etat k_1$ (see \Fig{pscaling}).
Finally, we recall that for the $\alpha^2$ dynamos considered here
both $\meanB_x$ and $\meanB_y$ oscillate, but their relative phase shift
is such that $\meanBB^2$ is non-oscillatory.
The normalized cycle frequency, $\tilde\omega\equiv\omega/\etat k_1$,
is also plotted in \Fig{pscaling} as a function of $\Rm$.
It is somewhat surprising that $\omega$ does not strongly depend on $\Rm$.
One may have expected that the cycle frequency could scale with the
inverse resistive time $\eta k_1^2$.
On the other hand, for oscillatory $\alpha\Omega$ dynamos the cycle
frequency is known to scale with $\etat k_1^2$ \cite{BB02}, although that
value could decrease if $\etat(\meanBB)$ is strongly quenched.
However, simulations only give evidence for mild quenching \citep{BRRS08,KB09}.

There is a dramatic difference between the cases with
and without magnetic helicity fluxes.
For $C_U=0.6$ the dynamo reaches asymptotic behavior
for large values of $\Rm$, while for $C_U=0$ the
saturation field strength goes to zero and $\max(-\alphaM)$
reaches quickly an asymptotic value corresponding to
a level of quenching that makes the dynamo marginally excited.

\begin{figure}\begin{center}
\includegraphics[width=\columnwidth]{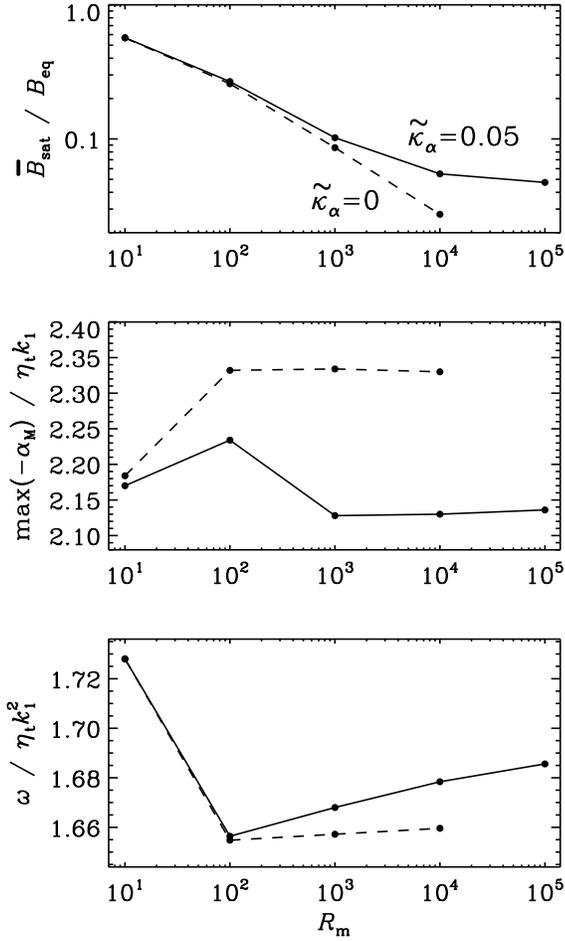}
\end{center}\caption[]{
Same as \Fig{pscaling}, but for antisymmetric solutions in a model
with perfect conductor boundary conditions with $C_U=0$ and
$\tilde\kappa_\alpha\equiv\kappa_\alpha=0.05$ (solid lines)
or 0 (dashed lines).
}\label{pscaling_equator}\end{figure}

\subsection{Helicity fluxes through the equator}

We have seen in \Sec{Kinematic} that in the perfect conductor case
the antisymmetric solutions are the most easily excited ones.
The boundary conditions for antisymmetric solutions permit magnetic
helicity transfer through the equator.
However, this alone does not suffice to alleviate catastrophic
quenching unless a sufficiently strong flux is driven through
the equator.
A possible candidate for driving such a flux
would be a diffusive flux driven by the $\nab\alphaM$ term.
In \Fig{pscaling_equator} we plot the $\Rm$ dependence of
$\max(-\alphaM)$, $\meanB_{\rm sat}$, and $\tilde\omega$
for $\tilde\kappa_\alpha=0.05$ and 0.
Again, catastrophic $\alpha$ quenching is alleviated by the
action of a magnetic helicity flux, but this time it is through the equator.
The maximum value of $-\alphaM$ is lowered by 15\% from 2.35 to 2.15
in units of $\etat k_1$ (see \Fig{pscaling_equator}).
Again, the cycle frequency is not changed significantly.

\begin{figure}\begin{center}
\includegraphics[width=\columnwidth]{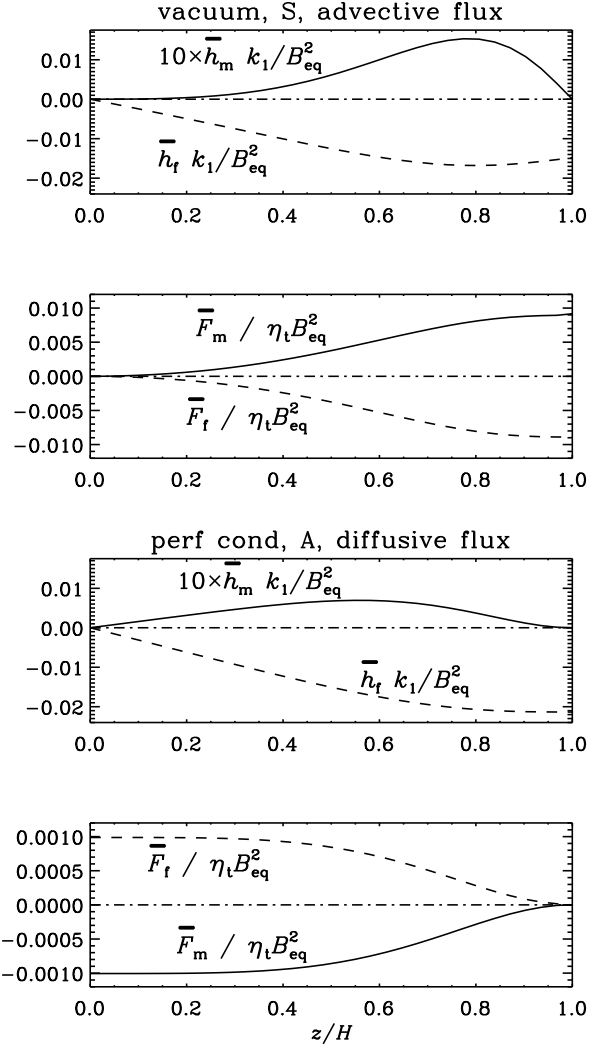}
\end{center}\caption[]{
Mean magnetic helicity densities of mean and fluctuating fields as
well as mean magnetic helicity fluxes of mean and fluctuating fields
as functions of $z$ for the S solution with vacuum boundary condition
and advective flux with $C_U=0.6$ (upper two panels) and for the A
solution with perfect conductor boundary condition and diffusive flux
with $\tilde\kappa_\alpha=0.05$ (lower two panels).
The profiles of $\meanhf$ have been scaled by a factor of 10 to
make them more clearly visible.
In all cases we used $C_\alpha=8$ and $\Rm=10^5$.
}\label{pflux_perfvert}\end{figure}

In \Fig{pflux_perfvert} we compare the profiles of $\meanhm$, $\meanhf$,
$\meanFm$, and $\meanFf$ for the most easily excited solution with vacuum
and perfect conductor boundary conditions on $z=L$.
In all cases we have $\meanhm=\meanhf=0$ at the midplane due to symmetry
reasons, and at $z=L$ we have $\meanhm=0$ and $\meanhf\neq0$.
It turns out that the magnetic helicity flux of the small-scale
field is balanced nearly exactly by that of the mean field.
This agrees with the expectation of \cite{BB03} who argued that
both should be shed at nearly the same rate.

The ad hoc assumption of a turbulent magnetic helicity flux is
plausible and has of course been made in the past \citep{KMRS03},
but its effect in alleviating catastrophic
quenching has not yet been seen in earlier three-dimensional turbulence
simulations \citep{BD01,B01b}.
However, except for the effects of boundaries, the conditions in those
simulation were essentially homogeneous and the gradients of magnetic
helicity density may have been just too small.
It would therefore be important to reconsider the question of diffusive
helicity fluxes in future simulations of inhomogeneous helical turbulence.

\subsection{Occasional removal of $\meanhf$}

Catastrophic quenching can also be alleviated by the artificial removal
of small-scale magnetic fields.
This was first demonstrated using three-dimensional simulations of
helical turbulence \citep{BDS02}.
In that study the small-scale magnetic field was isolated in Fourier
space and removed in regular time intervals.
Here we do the same by just resetting $\meanhf$ to a reduced value,
as described in \Eq{removal}.

\begin{figure}\begin{center}
\includegraphics[width=\columnwidth]{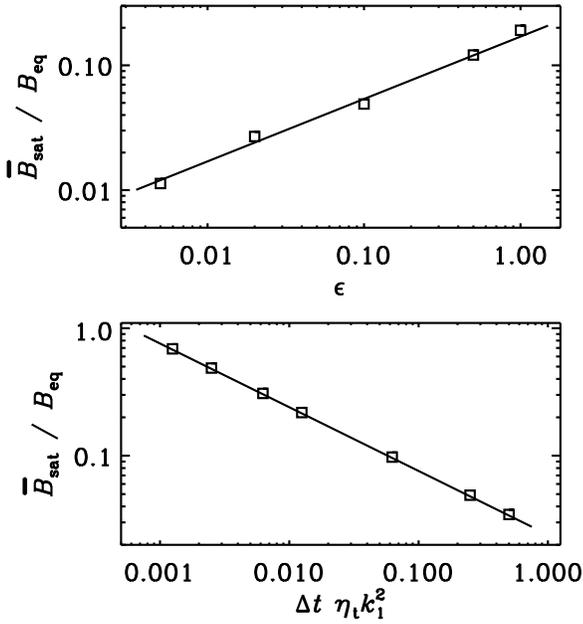}
\end{center}\caption[]{
Saturation field strength versus $\epsilon$ for $\Delta t\etat k_1^2=0.25$
(upper panel) and versus $\Delta t\etat k_1^2$ for $\epsilon=0.1$
(lower panel) in a model with $C_\alpha=8$, $\Rm=10^5$, and
$C_U=C_S=\kappa_\alpha=0$.
}\label{pscaling_eps_dt}\end{figure}

We consider the saturation strength of the magnetic field,
$\meanB_{\rm sat}$, to characterize the alleviating effect of
small-scale magnetic helicity losses.
Not surprisingly, the dynamo becomes stronger ($\meanB_{\rm sat}$
increases) when the fraction of small-scale field removal $\epsilon$
is increased (upper panel of \Fig{pscaling_eps_dt}) or the time interval
of field removal is decreased (lower panel of \Fig{pscaling_eps_dt}).
These dependencies follow approximate power laws,
\EQ
\meanB_{\rm sat}/\Beq^2
\approx0.17\,\epsilon^{1/2}
\approx0.024\,(\Delta t\etat k_1^2)^{-1/2},
\EN
suggesting that even relatively small amounts of magnetic helicity
removal in long intervals can have an effect.

We have also performed some numerical experiments where the magnetic helicity
associated with the small-scale field is only removed near the surface
layers.
However, in those cases the catastrophic quenching was not noticeably
alleviated.
This can be explained by noticing that, in the absence of additional
magnetic helicity fluxes in the interior, there is still a build-up
of $\meanhf$ in the interior which quenches the $\alpha$ effect
catastrophically.

\subsection{Magnetic helicity density and flux profiles}

\begin{figure}\begin{center}
\includegraphics[width=\columnwidth]{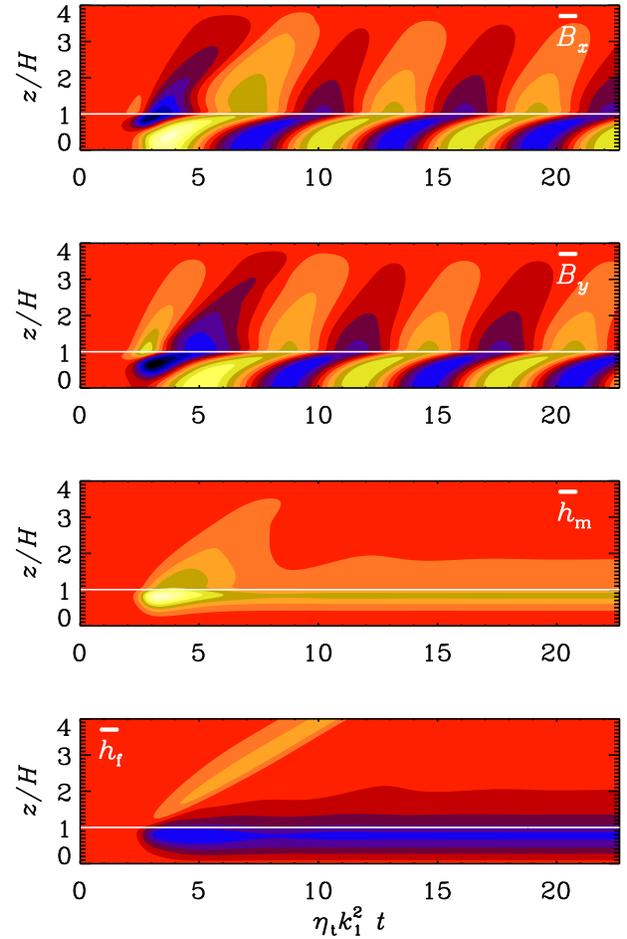}
\end{center}\caption[]{
Space-time diagrams for $\meanB_x$ and $\meanB_y$, as well as the magnetic
helicity densities $\meanhm$ and $\meanhf$ for $L=4H$, $C_\alpha=8$,
$C_U=0.6$, and $H_U\to\infty$.
The white horizontal line marks the location $z=H$.
Light (yellow) shades indicate positive values and dark (blue) shades
indicate negative values.
}\label{pbutter_hel_alpprof3_Uprof3_kappa0_Rm7}\end{figure}

\begin{figure}\begin{center}
\includegraphics[width=\columnwidth]{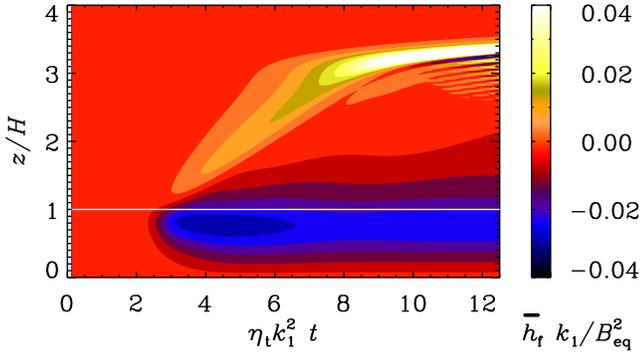}
\end{center}\caption[]{
Evolution of $\meanhf$ in a model similar to that of 
\Fig{pbutter_hel_alpprof3_Uprof3_kappa0_Rm7}, but for
$H_U\to\infty=3H$, so the flux of magnetic helicity
of the fluctuating field is blocked at $z=3H$.
Note the emergence of a shock that can eventually
no longer be resolved and leads to wiggles.
The white horizontal line marks the location $z=H$.
}\label{pbutter_hel_alpprof3_Uprof4_shock}\end{figure}

In an attempt to understand further the evolution of magnetic
helicity we have performed calculations where the magnetic helicity
flux of the fluctuating field was forced to vanish near the surface.
This was done by choosing a profile for $\meanU$ that goes to zero.
However, this invariably led to numerical problems.
In order to clarify the origin of these numerical problems
we chose to adopt a taller domain $L=4H$ using the profiles shown
in \Fig{pprof}.
The result is shown in \Fig{pbutter_hel_alpprof3_Uprof3_kappa0_Rm7}.
In that case the flux is still able to carry magnetic helicity away
from the dynamo-active region into the outer layers $z>H$.
The cyclic dynamo in $0\leq z\leq H$ operates otherwise very much
like before in the kinematic regime (\Fig{pbutter_kin}).
Dynamo action is possible for $C_\alpha>C_\alpha^{\rm crit}=4.32$.
However, a problem still arises when a parcel of positive
magnetic helicity that is shed early on from the dynamo-active region
reaches the upper layers at $z\approx3H$, through which now no magnetic
helicity can be transmitted.
Positive magnetic helicity builds up near $z\approx3H$ until wiggles
appear (\Fig{pbutter_hel_alpprof3_Uprof4_shock}).
This demonstrates that, once a magnetic helicity flux
is initiated, there is no way to stop it locally.
Instead, wiggles develop, so the solution becomes numerically invalid
at that moment.

The fact that positive magnetic helicity is produced is somewhat
unexpected, because for $\alpha>0$ the magnetic helicity production
is positive definite.
However, this can be traced back to the term $\etat\meanJJ\cdot\meanBB$,
which is part of $\meanEMF\cdot\meanBB$ on the right hand side of \Eq{dhf}.
Since $\meanJJ\cdot\meanBB$ is positive for positive $\alphaK$, it is clear
that this term produces positive $\meanhf$ just {\it outside} the range where
$\alphaK$ is finite and where it would produce $\meanhf$ of opposite sign.

\begin{figure}\begin{center}
\includegraphics[width=\columnwidth]{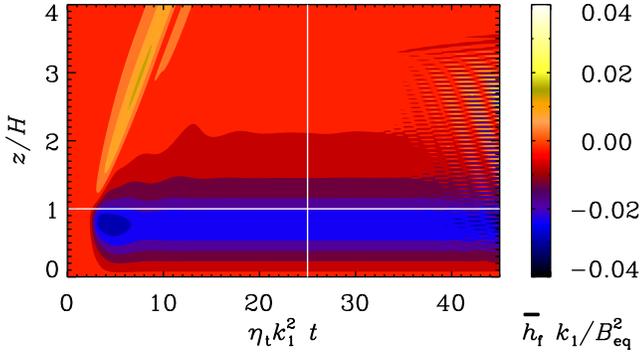}
\end{center}\caption[]{
Similar to \Fig{pbutter_hel_alpprof3_Uprof4_shock}, but this time
the value of $H$ has been changed from infinity to $H_U=3H$ only
at the time $\etat k_1^2t=25$ (marked by a vertical white line).
Soon after that time the solution develops wiggles that invalidate
the results after that point.
}\label{pbutter_hel_alpprof3_Uprof4_shock2}\end{figure}

In another experiment we adopt a profile for $\meanUU$ such that $H_U$
is changed from $\infty$ to $3H$ only after a time $t\etat k_1^2=25$,
which is when the positive bump of $\meanhf$ has left the domain.
The result is shown in \Fig{pbutter_hel_alpprof3_Uprof4_shock2}.
Now it is indeed negative magnetic helicity that the dynamo tries to shed.
However, even though the flux is relatively weak, the blockage at $z=3H$
leads eventually to a problem and leads, again, to wiggles indicating
that the solution is numerically not valid.

\begin{figure}\begin{center}
\includegraphics[width=\columnwidth]{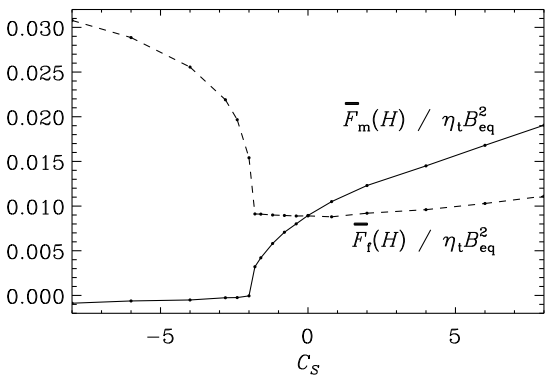}
\end{center}\caption[]{
Dependence of $\meanFm(H)$ and $\meanFf(H)$ on the shear parameter
for the S solution in a model with vacuum boundary condition,
$C_\alpha=8$, $C_U=0.6$, and $\Rm=10^5$.
}\label{pFmFf_vs_shear}\end{figure}

These results suggest that the magnetic helicity flux must be allowed
to continue through the rest of the domain.
Of course, in reality there is the possibility of various fluxes,
including diffusive fluxes that have not been included in this
particular model.
We note, however, that model calculations with finite $\kappa_\alpha$
in \Eq{FluxFormula} confirm that then $\meanFf(z,t)$ becomes constant
in the outer parts.

\subsection{Magnetic helicity with shear}

It is remarkable that the magnetic helicity fluxes of the mean and
fluctuating fields were always equally strong and of opposite sign.
The point of this section is to underline that this is a particular
property of the $\alpha^2$ dynamo, and would not carry over to
$\alpha\Omega$ dynamos.
In \Fig{pFmFf_vs_shear} we show the fluxes of the model with
$C_\alpha=8$ and $C_U=0.6$ where we have varied $C_S$ in the range
from $-8$ to $+8$.

It turns out that the presence of shear gives rise to an additional
magnetic helicity flux \citep{BR00} and that the perfect correspondence
between magnetic helicity fluxes of opposite sign for mean and fluctuating
fields is then broken.
This additional flux of magnetic helicity is associated with the mean
field and does therefore not, on its own, alleviate catastrophic quenching.
However, in this model we have neglected magnetic helicity fluxes that
would be associated with the fluctuating field.
An example is the Vishniac-Cho flux whose effect in a mean-field model
was already studied in an earlier paper \citep{BS05}.
For $C<-2$, the oscillating solutions are no longer preferred and a new
solution branch emerges, where the solutions are now non-oscillatory.
Those are also the type of solutions studied by \cite{SSSB06},
where $C_S=-8$ was chosen, corresponding to the value $-2$ in their
normalization.

\section{Conclusions}

The present simulations have confirmed that in finite domains magnetic
helicity losses through local fluxes are able to alleviate catastrophic
quenching.
Without such fluxes the energy of the mean field goes to zero in the
limit of large $\Rm$, while in the presence of such fluxes $|\meanBB|$
reaches values that are about $5\%$ of the equipartition value.
We emphasize at this point that this applies to the case of an $\alpha^2$
dynamo.
For an $\alpha\Omega$ dynamo the mean field can reach larger values,
depending on the amount of shear.
For example for the model shown in \Fig{pFmFf_vs_shear} the field
strength in units of the equipartition value rises from 5\% without
shear to about 36\% with negative shear ($C_S=-8$), while
for positive shear it stays around 5\%.
We also emphasize that the difference between the two cases with and
without helicity fluxes is rather weak for $\Rm\leq10^3$, so one
really has to reach values around $\Rm\leq10^4$ or $\Rm\leq10^5$ to
make a clear difference compared with catastrophic quenching.
Such high values of $\Rm$ are not currently feasible with three-dimensional
turbulence simulations.

The other surprising result is that it is not possible
to dissipate magnetic helicity flux locally once it is initiated.
If the magnetic helicity flux of the small-scale field has already left
the dynamo-active domain, it has to stay constant in the steady state.
By adding a diffusive flux the shock-like structure in the magnetic
helicity of the small-scale field could be smoothed out, but this
contribution would then carry the same amount of energy as before,
although now by other means.

In the presence of shear there are additional contributions to the
magnetic helicity flux associated with the mean magnetic field.
There are first of all the fluxes associated with the mean field
itself, but those fluxes cannot contribute to alleviating catastrophic
quenching on their own.
However, earlier work has shown that in the presence of shear there
are also additional contributions associated with the fluctuating
field \citep{VC01,SB04,SB06}.
Those terms have not been included in the present work, because
they have already been studied in an earlier paper \citep{BS05}.

Several new issues have emerged from the present study.
The fact that diffusive magnetic helicity fluxes through the equator
can alleviate catastrophic quenching is not surprising as such, but
its effects in alleviating catastrophic saturation behavior in
three-dimensional turbulence simulations has not yet been reported
\citep{BD01,B01b}.
On the other hand, simulations of forced turbulence in spherical
shells with an equator did show near-equipartition strength
saturations fields \citep{Dhruba}, although the values of $\Rm$
were typically below 20, so it was not possible to draw conclusions
about catastrophic quenching.
A new dedicated attempt in that direction would be worthwhile using
again driven turbulence, but now with a linear gradient of its intensity
and in cartesian geometry.

In view of applications to the Sun and other stars,
another important development would be to extend the present work to
spherical domains.
Again, some work in that direction was already reported in
\cite{BKMMT07}, but none of these models used diffusive fluxes, nor
has any attempt been made to model the Sun.
This would now be an important target for future research.

\section*{Acknowledgments}
We acknowledge the use of computing time at the Center for
Parallel Computers at the Royal Institute of Technology in Sweden.
This work was supported in part by
the European Research Council under the AstroDyn Research Project 227952
and the Swedish Research Council grant 621-2007-4064.


\vfill\bigskip\noindent\tiny\begin{verbatim}
$Header: /var/cvs/brandenb/tex/simon/1d_helicity_qnch/paper.tex,v 1.36 2009-05-03 19:52:37 brandenb Exp $
\end{verbatim}

\end{document}